\documentclass[twocolumn,preprintnumbers,superscriptaddress,amsmath,amssymb]{revtex4}
\usepackage{float}
\usepackage{graphicx}
\usepackage{dcolumn}
\usepackage{bm}
\usepackage{epstopdf}
\usepackage{xcolor}
\usepackage{textcomp}
\usepackage{soul}
\usepackage{csquotes}
\usepackage{amsbsy}
\usepackage{mathrsfs} 
\usepackage{amsmath}
\usepackage{bigints} 
\usepackage{amsthm,amssymb} 
\usepackage[utf8]{inputenc}
\usepackage[english]{babel}
\usepackage[shortlabels]{enumitem}
\usepackage{float}
\usepackage{mathrsfs,bigints,mathtools,dsfont}
\usepackage[colorlinks=true,linkcolor=blue,citecolor=blue]{hyperref}%
\usepackage[toc]{appendix}
\newcommand\norm[1]{\left\lVert#1\right\rVert}

\begin{document}
	
\title{{Stability of synchronization in simplicial complexes with multiple interaction layers}}
\author{Md Sayeed Anwar}\affiliation{Physics and Applied Mathematics Unit, Indian Statistical Institute, 203 B. T. Road, Kolkata 700108, India}
\author{Dibakar Ghosh}\email{dibakar@isical.ac.in}\affiliation{Physics and Applied Mathematics Unit, Indian Statistical Institute, 203 B. T. Road, Kolkata 700108, India}

\begin{abstract}
Understanding how the interplay between higher-order and multilayer structures of interconnections influences the synchronization behaviors of dynamical systems is a feasible  problem of interest, with possible application in essential topics such as neuronal dynamics. Here, we provide a comprehensive approach for analyzing the stability of the complete synchronization state in simplicial complexes with numerous interaction layers. We show that the synchronization state exists as an invariant solution and derive the necessary condition for a stable synchronization state in presence of general coupling functions. It generalizes the well-known master stability function scheme to the higher-order structures with multiple interaction layers. We verify our theoretical results by employing them  on networks of paradigmatic R\"{o}ssler oscillators and Sherman neuronal models, and demonstrate that the presence of group interactions considerably improves the synchronization phenomenon in the multilayer framework.   	
\end{abstract}	
		
\maketitle

\section{Introduction}
Network theory \cite{boccaletti2006complex,newman2003structure} has proven to be an excellent foundation for modeling a variety of natural and artificial systems. One of the most appealing aspects of this approach is its ability to recognize unified characteristics in the framework of connections between fundamental elements of a system \cite{watts1998collective,dorogovtsev2008critical}. The study of dynamical systems on networks has thus piqued scientists' curiosity, with applications ranging from engineering and physics to ecology and social science \cite{newman2018networks,latora2017complex}. These network models are premised on the hypothesis that the components of a system are connected only through pairwise links, schematized by graphs. However, this pairwise representation is merely a first-order estimation \cite{beyond_pairwise,majhi2022dynamics} in many real-world systems, such as structural \cite{structural_brain} and functional \cite{functional_brain1} brain networks, ecological networks \cite{grilli2017higher,billick1994higher}, collaboration graphs \cite{coauthor1,coauthor2}, protein interaction networks \cite{protein}, social systems \cite{benson2016higher,random_pre}, and consensus dynamics \cite{schaub2}, where many-body interactions between system components are ubiquitous and of immense interest. To schematize these higher-order interactions, one then requires a generalization of classical network theory \cite{bick2021higher,battiston2021physics}.  
\par Simplicial complexes \cite{torres2020simplicial,giusti2016two}, a topological structure formed by an ensemble of simplices of different dimensions, are a natural generalization of graphs to describe these group interactions. A $d$-simplex is a set of $(d+1)$ nodes $\sigma=[v_{0}, v_{1}, \cdots, v_{d}]$, namely the $0$-simplices are the vertices, the edges are $1$-simplices, while $2$-simplices correspond to full triangles and so on. A simplicial complex $\mathscr{S}$ is a collection of simplices with an extra constraint that if simplex $\sigma \in \mathscr{S}$ then all the subsets of $\sigma$ also belong to $\mathscr{S}$, i.e., a three-body interaction, for example, necessitates the presence of all pairwise connections corresponding to the same triangle. A simplicial complex is called $D$-dimensional if the maximal simplices contained in it are of dimension $D$. Although the concept of simplicial complexes is not new \cite{aleksandrov1998combinatorial,hypergraph1}, the recent advances in real-world data availability \cite{carlsson2009topology} have renewed the interest of researchers in examining the collective dynamical behavior in simplicial complexes, which shows some interesting results due to the inclusion of many-body interactions among dynamical entities \cite{contagion1,ecological2,simplicialsync4,beyond_pairwise,simplicial3}. 
\par Synchronization \cite{synchronization1,synchronization2,synchronization3}, or the formation of the order in ensembles of interacting individuals, is one such fascinating emergent phenomenon where the connection between isolated dynamical elements plays an important role. Researchers have recently expanded the study of synchronization to the framework including higher-order network structures, with the majority of them opting for simplicial complexes to simulate group interactions due to their simple topological representation \cite{simplicial2,simplicialsync3}. The presence of many-body interactions has been linked to the emergence of abrupt synchronization transitions \cite{simplicialsync4,skardal2020higher,kuehn2021universal,kachhvah2022hebbian}, improvement of synchronization \cite{skardal2021higher,zhang2022higher}, multistability  \cite{xu2020bifurcation}, cluster synchronization \cite{zhang2021unified}, antiphase synchronization \cite{kachhvah2022first}, chimeras \cite{srilena_chimera}, etc. These findings have coincided with the development of analytical paradigms for interpreting coupled oscillators with many-body interactions, such as Hodge decomposition \cite{millan2020explosive,arnaudon2021connecting}, Laplacian operators \cite{simplicialsync2,simplicialsync6}, and  low dimensional descriptions \cite{gong2019low}. 
\par On the other hand, multilayer systems are other generalized network structures in which interactions of different kinds and meanings can coexist, resulting in networks of networks \cite{sorrentino2012synchronization,del2016synchronization}. In these systems, nodes can be linked using various types of connections, and each of these connection types characterizes an interaction layer \cite{boccaletti2014structure}. Mobility networks, where different modes of transportation connect individual elements \cite{mobility}, social networks, wherein individuals are interconnected and acquainted by various kinds of relations \cite{social}, and neuronal networks, where the neurons communicate through chemical and electrical channels \cite{rakshit2018synchronization}, are examples of multilayered systems. Various collective phenomena, such as synchronization \cite{intra2,inter2}, percolation \cite{bianconi2014multiple}, diffusion \cite{gomez2013diffusion}, epidemic spreading \cite{granell2013dynamical}, and evolutionary games \cite{gomez2012evolution}, have been investigated in multilayer frameworks, all revealing a phenomenology that differs significantly from that observed in monolayer structures. In all these studies, the connections between individual nodes of a layer are considered to be pairwise, schematized by links. However, this pairwise connectivity representation within layers may not always be able to illustrate the connection topology accurately among individual nodes. The neuronal system is an example where, on the one hand, individual neurons are connected via electrical and chemical synapses \cite{rakshit2018synchronization,parastesh2022}, and on the other hand, proof of many-body interactions between individual neurons have recently been found \cite{beyond_pairwise,ince2009presence,tlaie2019high,amari2003synchronous}. Nevertheless, the interplay between higher-order structures and multilayer networks \cite{sun2021higher}, under some aspects has yet to be investigated, and specifically, the study of synchronization is still in its early stages \cite{anwar2022intralayer,jalan2022multiple}.  In the previous study \cite{anwar2022intralayer} of synchronization on multilayer framework with higher-order interactions, the interactions between individual elements of a particular layer are considered to be of a specific functional form (linear diffusive). These linear interactions factorize the many-body structures into pairwise interactions due to the superposition property of linear functions, resulting in a weighted pairwise network. Nevertheless, this weighted pairwise network representation cannot always capture the group interactions between individuals, especially when the interactions between unitary elements are nonlinear \cite{beyond_pairwise,schaub2} ( e.g., in neuronal systems). The higher-order interactions cannot be factorized into weighted pairwise connections in such circumstances, and therefore there is still room for improvement in the investigation of synchronization on higher-order multilayered systems. 
\par In this work, we fill this gap by investigating the interplay between multilayer and higher-order structures and providing a generic multilayer framework to study the synchronization phenomenon on simplicial complexes. Our goal is to develop a comprehensive mathematical approach for evaluating the stability of the synchronization state for nonlinear dynamical systems growing in simplicial complexes with numerous interaction layers. To do so, we first derive the condition of invariance for the synchronization state and then conduct the linear stability analysis of the synchronization solution, which as a result generalizes the well-known master stability function (MSF) approach \cite{msf} to the realm of simplicial complexes \cite{simplicialsync2} with multiple interaction layers \cite{del2016synchronization}. The end result is a system of coupled linear equations ( master stability equation ) for the evolution of separation of the simplicial complex, which can be optimally separated for a particular instance. The proposed approach does not require any restriction on the functional forms of the coupling schemes between individual elements of the layers. We validate all our theoretical predictions in a simplicial complex of coupled Sherman neuronal model and chaotic R\"{o}ssler oscillators with two layers of interactions featuring different nonlinear coupling schemes and connection topologies. 
\par The remaining of this article is structured in the following manner. A mathematical framework for the simplicial complexes with multiple interaction layers is proposed in Sec. \ref{mathematical_framework}. Section \ref{Theoretical treatments} shows the theoretical results on the stability of the synchronization state. The invariance criterion for the synchronization state is obtained in Sec. \ref{invariance}. Following that, in Sec. \ref{stability}, we develop the analytical condition for the stable synchronous solution. To validate our theoretical findings, in Sec. \ref{numericals}, we reported numerical results on systems of coupled oscillators. Finally, in Sec. \ref{conclusion}, we summarize all of our findings and draw a conclusion.              
\section{Generalized Mathematical Model of the simplicial complex} \label{mathematical_framework}
We consider a simplicial complex of dimension $D=2$, i.e., the interactions between dynamical units are not only limited to pairwise but also three-body interactions are allowed. The simplicial complex is composed of $N$ dynamical units connected through $M$ distinct interacting layers. It is worth noting that the nodes communicating on each layer in this setup are essentially the same nodes. The node $i$ in the layer- $1$ is identical to the node $i$ in all the other layers.  Various sorts of transport between cities or electrical and chemical synaptic couplings between neurons are examples of complex systems with multiple interaction layers where essentially the same nodes are interconnected by different means of connections. This differs from other works \cite{rakshit2018synchronization,intra2,relay2,anwar2022intralayer,inter2} where nodes in various layers  can be identical or different from each other, and further, the nodes in adjacent layers are connected through a set of links called the interlayer connections. Now, we further consider that among all the layers, within $L$ $(1 \leq L \leq M)$ number of layers a $2$-simplex $(i,j,k)$ is created by promoting an empty triangle composed of three $1$-simplices $(i, j)$, $(j, k)$, $(k, i)$ to a full triangle $(i, j, k)$, i.e., all the cliques of size $3$ are considered as a $2$-simplex. Therefore, these $L$ layers consider both the two-body and three-body interactions. However, in the other $(M-L)$ layers, no empty triangle composed of three $1$-simplices is promoted to a full triangle ($2$- simplex), i.e., these $(M-L)$ layers consider only the pairwise interactions between dynamical units. This implies that the proposed simplicial complex allows the presence of both empty and full triangles simultaneously, which is a more realistic model of simplicial complexes as in many real-world systems, the presence of pairwise interactions between three individuals does not necessarily imply a three-body interaction between them. It is important to note that the proposed model can easily be extended to any arbitrary higher-dimensional simplicial complexes (i.e., $D>2$), but to keep things simple, we consider only up to $D=2$ dimension. Further, all the derived theoretical results can be expanded to simplicial complexes of any dimension $D$ without any difficulty.
\begin{figure}[ht] 
	\centerline{
		\includegraphics[scale=0.25]{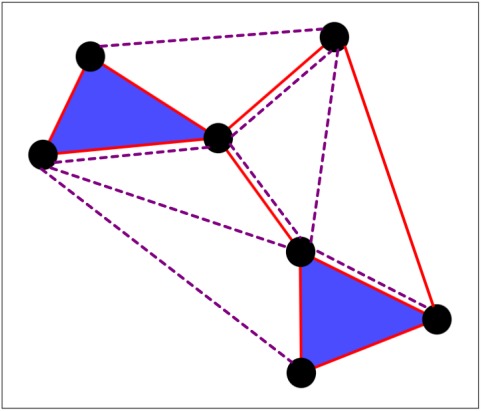}}
	\caption{ {\bf Diagrammatic representation of simplicial complex with two distinct interaction layers.} The two layers (solid red and dashed magenta links, respectively) are built up of different sorts of interconnections for the same nodes, where the layer with red links allows three-body interactions between the nodes. The layers are completely independent; the presence of a link between two nodes in one layer has no influence on their connection status in the other. }
	\label{schematic}
\end{figure}       
\par
We suppose that the equation of motion characterizing the simplicial complex dynamics can be stated in terms of the following equation,     
	\begin{equation}\label{gen_model}
		\begin{array}{lcl}
			\dot{\bf x}_i(t) = F({\bf x}_i)+\sum\limits_{\beta=1}^{M}\epsilon_{\beta}\sum\limits_{j=1}^{N}{\mathscr{A}_{ij}^{[\beta]}} G_{\beta}({\bf x}_i,{\bf x}_j) \\
			+ \sum_{q=1}^{L}\epsilon_{SI}^{q}\sum\limits_{j=1}^{N}\sum\limits_{k=1}^{N}{\mathscr{A}_{ijk}^{[q,SI]}} G_{SI}^{q}({\bf x}_i,{\bf x}_j,{\bf x}_k), \hspace{10 pt} i=1,2,...,N.
		\end{array}
	\end{equation}	
Here the $p$-dimensional state vector $\mathbf{x}_{i}(t)$ indicates the dynamics of the unit $i$, $F : \mathbb{R}^p \rightarrow  \mathbb{R}^p$ illustrates the isolated node dynamics presumed to be identical for all units, $\epsilon_{\beta}$ $(\beta=1,2,\cdots,M)$ are the real-valued coupling strengths corresponding to the pairwise interactions within the layers, whereas $\epsilon_{SI}^{q}$ describes the coupling strength associated with the three-body ($2$-simplex) interactions among dynamical units of the layer where higher-order interactions are allowed. $G_{\beta}:  \mathbb{R}^{(2 \times p)} \rightarrow  \mathbb{R}^p$, $G_{SI}^{q}:  \mathbb{R}^{(3 \times p)} \rightarrow  \mathbb{R}^p$ are the real-valued continuously differentiable functions illustrating the coupling forms corresponding to pairwise and non-pairwise interactions, respectively. Here we can consider the coupling functions in arbitrary linear or nonlinear form. Furthermore, the $N \times N $ adjacency matrices $\mathscr{A}^{[\beta]}$ $(\beta=1,2,\cdots,M)$ recount the network structures of the layers, where $\mathscr{A}^{[\beta]}_{ij}=1$, if the $i$-th and $j$-th units are interconnected and zero otherwise. $\mathscr{A}^{[q,SI]}$ describes $N \times N \times N$ adjacency tensors that account for $2$-simplices, where $\mathscr{A}^{[q,SI]}_{ijk}=1$, if $(i,j,k)$ construct a full triangle and zero otherwise. If one considers $M=1$ and $L=1$, then our mathematical framework will eventually coincide with the generalized framework proposed earlier in \cite{simplicialsync2} for mono-layer representation of simplicial complexes. Therefore, Eq. \eqref{gen_model} represents a generalized mathematical framework to investigate dynamical processes in simplicial complexes with multiple interaction layers.		
 \par A schematic illustration of the proposed simplicial complex with $N=7$ nodes and $M=2$ distinct interaction layers is shown in Fig. \ref{schematic}. The connection mechanisms between the nodes via two different layers are represented in two different colors; solid red lines correspond to the first layer and dotted magenta lines for the second layer. In the first layer, both pairwise and three-body interactions are considered, where all the three cliques (empty triangles) are promoted to full triangles ($2$-simplices), shaded in blue color. However, in the second layer, no empty triangles are promoted to full triangles.     

\section{Theoretical approaches} \label{Theoretical treatments}
In this section, our main goal is to analytically derive the necessary conditions for the  stable complete synchronization state of the simplicial complex \eqref{gen_model}. To do so, we first look into the conditions for invariance of the synchronization state. The invariance condition gives the necessary requirement on the network topology by which a stable synchronization state may observe depending on the critical value of the coupling strengths.
\subsection{Invariance of synchronization state }\label{invariance}
\par A complete synchronization state arises in the simplicial complex \eqref{gen_model} when each node follows in unison with the rest nodes. Mathematically, there exists a solution $\mathbf{x}_0(t) \in \mathbb{R}^p$ such that, 
\begin{equation*}
	\begin{array}{l}
	\mbox{for}~~ t \to \infty,	~~ \norm{\mathbf{x}_i(t)-\mathbf{x}_0(t)} \to 0, ~~ i=1,2,\dots,N.
	\end{array}		
\end{equation*} 
Following this, the corresponding synchronized manifold is defined as follows, 
\begin{equation*}\label{eq.}
	\begin{array}{l}
		\mathcal{S}=\{\mathbf{x}_0(t) \in \mathbb{R}^{p} : \mathbf{x}_i(t)=\mathbf{x}_0(t),\\ ~~ \mbox{for} ~~ i=1,2,\dots,N ~~ \mbox{and} ~~ t \in \mathbb{R}^+ \}.	
	\end{array} 
\end{equation*} 
If the functional forms of coupling schemes are synchronization noninvasive \cite{simplicialsync2}, i.e.,  
\begin{equation} \label{noninvasive}
	\begin{array}{l}
		G_{\beta}({\bf x}_0,{\bf x}_0)=0, \; \mbox{and}\; G_{SI}^{q}({\bf x}_0,{\bf x}_0,{\bf x}_0)=0,	
	\end{array}
\end{equation} 
then the existence and invariance of the synchronous solution will be assured. However, we cast away any restriction on functional forms of coupling mechanisms, consider arbitrary linear or nonlinear coupling functions, and derive the invariance condition for such board circumstances. 
\par If the simplicial complex starts evolving with the synchronization state at any instance of time $t = t_s$, then $\mathbf{x}_i(t_{s})=\mathbf{x}_0{(t_{s})}$, for $i=1,2,\dots,N$. The evolution of the $i^{\mbox{th}}$ node of the simplicial complex at $t = t_s$ can then be expressed from Eq. \eqref{gen_model} as follows, 
\begin{equation}\label{invariance_1}
	\begin{array}{l}
		\dot{\bf x}_i(t_{s}) = F({\bf x}_0)+\sum_{\beta=1}^{M}\epsilon_{\beta} k_{(in)_i}^{\beta} G_{\beta}({\bf x}_0,{\bf x}_0) \\\\ ~~~~~~~~~~~~~~ + 2 \sum_{q=1}^{L}\epsilon_{SI}^{q} k_{(in)_i}^{[q,SI]} G_{SI}^{q}({\bf x}_0,{\bf x}_0,{\bf x}_0),
	\end{array}
\end{equation}
where $k_{(in)_i}^{\beta}= {\sum\limits_{j=1}^{N}{\mathscr{A}_{ij}^{[\beta]}}}$ denotes the number of links incident to node $i$ in each individual layer $\beta$, i.e., in-degree of node $i$ and $k_{(in)_i}^{[q,SI]}= \frac{1}{2}{\sum\limits_{j=1}^{N}\sum\limits_{k=1}^{N}{\mathscr{A}_{ijk}^{[q,SI]}}}$ indicates the number of $2$-simplices (full triangle) incident to node $i$, i.e., generalized in-degree of node $i$ \cite{directed_higher}. Now, in order to maintain the synchronous solution, each node must progress at the same rate. As a result, any two distinct nodes $i$ and $j$ in the simplicial complex should have the same velocity, i.e.,
\begin{equation}\label{equal_velocity}
	\begin{array}{l}
		\dot{\mathbf{x}}_i(t_{s}) = \dot{\mathbf{x}}_j(t_{s}), \;\; {\mbox{for any}~i\ne j,~\mbox{and}}~i,j=1,2,\dots,N.	
	\end{array}
\end{equation}
Since the functional forms of coupling schemes are arbitrary linear or nonlinear, Eq. \eqref{equal_velocity} together with Eq. \eqref{invariance_1} gives the following relation,
\begin{equation} \label{invariance_2}
	\begin{array}{l}
		k_{(in)_i}^{\beta}=k_{(in)_j}^{\beta}=k_{(in)}^{\beta} \;\mbox{(say)}, \; \beta=1,2,\cdots,M \\ 
		~~~~~~~ \mbox{and} \;k_{(in)_i}^{[q,SI]}=k_{(in)_j}^{[q,SI]}=k_{(in)}^{[q,SI]}  \;\mbox{(say)}.
	\end{array}
\end{equation}
{\it Hence for the consistency of complete synchronization state, the same number of edges must be incident to each node of layer-$\beta$ and for the layers with three-body interactions, exactly the same number of $2$-simplices must incident to each node.} However, we here stick ourselves with only the undirected connection structure of the simplicial complex. Therefore, in this case, $k_{(in)}^{\beta}=k^{\beta}$, the degree of each node and $k_{(in)}^{[q,SI]}=k^{[q,SI]}$, the generalized degree of each node.   
\par As a result, the synchronous solution evolves according to the following equation,
\begin{equation}\label{invariance_3}
	\begin{array}{l}
		\dot{\bf x}_0(t_{s}) = F({\bf x}_0)+\sum_{\beta=1}^{M}\epsilon_{\beta} k^{\beta} G_{\beta}({\bf x}_0,{\bf x}_0)
		\\ ~~~~~~~~~~~~~ + 2 \sum_{q=1}^{L} \epsilon_{SI}^{q} k^{[q,SI]} G_{SI}^{q}({\bf x}_0,{\bf x}_0,{\bf x}_0).
	\end{array}
\end{equation}
 At this point, we would like to emphasize that the derived invariance condition is only necessary for the layers with arbitrary linear or nonlinear coupling schemes between individual units. If the forms of interaction functions within any layer are synchronization noninvasive (i.e., satisfies Eq.\eqref{noninvasive}), then the invariance condition (Eq. \eqref{invariance_2}) is not necessary as the coupling terms vanish at the synchronization solution, and the existence of synchronous state can be guaranteed.
\par {\it Therefore, the invariance of the complete synchronization state depends either upon the synchronization noninvasive form of coupling schemes within the layers or upon the regular topology of the layers (i.e., equal in-degrees of the nodes in a particular layer).} However, to investigate the stability of the synchronization solution, we proceed with the case of arbitrary coupling schemes. The study corresponding to synchronization noninvasive coupling forms is just a special case of this, as the last two terms of Eq. \eqref{invariance_3} vanish for noninvasive couplings.    
\subsection{Linear stability analysis} \label{stability}
To derive the conditions for stable synchronization state, one considers a small perturbation $\delta \mathbf{x}_{i}= \mathbf{x}_{i}-\mathbf{x}_{0}$ around the synchronous solution $\mathbf{x}_{i}=\mathbf{x}_{0} ~ (i=1, 2, \cdots, N)$ and performs linear stability analysis of Eq. \eqref{gen_model}. This gives  
\begin{widetext}
	\begin{equation} \label{stability_2}
		\begin{array}{l}
			\delta \dot{{\bf x}}_{i}= JF({\bf x}_{0}) \delta{\bf x}_{i} + \sum_{\beta=1}^{M}\epsilon_{\beta} \sum_{j=1}^{N} \mathscr{A}^{[\beta]}_{ij} \bigg[J_{1}G_{\beta}{({\bf x}_{0},{\bf x}_{0})} \delta {\bf x}_{i}+ J_{2}G_{\beta}{({\bf x}_{0},{\bf x}_{0})} \delta {\bf x}_{j} \bigg]  \\
			
			\;\;\;\;\;\;\;\;\; +  \sum_{q=1}^{L}\epsilon_{SI} \sum_{j=1}^{N} \sum_{k=1}^{N} \mathscr{A}^{[q,SI]}_{ijk} \bigg[J_{1}G^{q}_{SI}{({\bf x}_{0},{\bf x}_{0},{\bf x}_{0})} \delta {\bf x}_{i}+ J_{2}G^{q}_{SI}{({\bf x}_{0},{\bf x}_{0},{\bf x}_{0})} \delta {\bf x}_{j} + J_{3} G^{q}_{SI}{({\bf x}_{0},{\bf x}_{0},{\bf x}_{0})} \delta {\bf x}_{k} \bigg],
		\end{array}
	\end{equation}
\end{widetext}
where $JF({\bf x}_{0})$ indicates the Jacobian of $F$ calculated at the synchronization solution $\mathbf{x}_{0}$. $ J_{1}$, $J_{2}$, and $J_{3}$ are the Jacobian operators with respect to the first, second and third variables, respectively. $\mathscr{L}^{[\beta]}$, $\beta=1,2,\cdots,M$ are standard zero row-sum graph Laplacian matrices, defined by 
\begin{equation*}
	\begin{array}{l}
		\mathscr{L}^{[\beta]}_{ij}= \begin{cases}
			-\mathscr{A}^{[\beta]}_{ij}, & i \neq j \\
			k^{\beta}_{i}, & i=j.
		\end{cases}
	\end{array}
\end{equation*}
Now one has that  $\sum_{j=1}^{N} \mathscr{A}^{[\beta]}_{ij}= k_{i}^{\beta}$, $\sum_{j=1}^{N} \sum_{k=1}^{N} \mathscr{A}^{[q,SI]}_{ijk}=2k_{i}^{[q,SI]}$ and the invariance condition \eqref{invariance_2}. Hence plugging back the terms in the Eq. \eqref{stability_2}, one eventually obtains,
\begin{widetext}
	\begin{equation} \label{stability_3}
		\begin{array}{l}
			\delta \dot{{\bf x}}_{i}= JF({\bf x}_{0}) \delta{\bf x}_{i} + \sum_{\beta=1}^{M}\epsilon_{\beta} k^{\beta} \bigg[J_{1}G_{\beta}{({\bf x}_{0},{\bf x}_{0})} + J_{2}G_{\beta}{({\bf x}_{0},{\bf x}_{0})} \bigg] \delta {\bf x}_{i} - \sum_{\beta=1}^{M}\epsilon_{\beta} \sum_{j=1}^{N} \mathscr{L}^{[\beta]}_{ij} J_{2}G_{\beta}{({\bf x}_{0},{\bf x}_{0})} \delta {\bf x}_{j}  \\
			
			\;\;\;\;\;\; + 2 \sum_{q=1}^{L}\epsilon^{q}_{SI} k^{[q,SI]} J_{1}G^{q}_{SI}{({\bf x}_{0},{\bf x}_{0},{\bf x}_{0})} \delta {\bf x}_{i} + \sum_{q=1}^{L}\epsilon^{q}_{SI} \sum_{j=1}^{N} \sum_{k=1}^{N} \mathscr{A}^{[q,SI]}_{ijk} \bigg[ J_{2}G^{q}_{SI}{({\bf x}_{0},{\bf x}_{0},{\bf x}_{0})} \delta {\bf x}_{j} + J_{3} G^{q}_{SI}{({\bf x}_{0},{\bf x}_{0},{\bf x}_{0})} \delta {\bf x}_{k} \bigg].
		\end{array}
	\end{equation}
\end{widetext}
At this stage of derivation, it is important to note that our approach does not require any specific functional form of coupling functions, for instance, diffusive or synchronization noninvasive \cite{simplicialsync2} functional forms to derive the stability condition of the synchronization state, and as a result, we are literally encompassing arbitrary sorts of coupling functions. The derivation for diffusive or synchronization noninvasive functional forms can be considered as some special cases of our approach. 
\par Let us now use the fact that the Jacobians $J_{2}G^{q}_{SI}{({\bf x}_{0},{\bf x}_{0},{\bf x}_{0})}$, $J_{3} G^{q}_{SI}{({\bf x}_{0},{\bf x}_{0},{\bf x}_{0})}$ are independent of $k$ and $j$, respectively. This simplifies the Eq. \eqref{stability_3} as,
\begin{widetext}              
	\begin{equation} \label{stability_4}
		\begin{array}{l}
			\delta \dot{{\bf x}}_{i}= JF({\bf x}_{0}) \delta{\bf x}_{i} + \sum_{\beta=1}^{M}\epsilon_{\beta} k^{\beta} \bigg[J_{1}G_{\beta}{({\bf x}_{0},{\bf x}_{0})} + J_{2}G_{\beta}{({\bf x}_{0},{\bf x}_{0})} \bigg] \delta {\bf x}_{i} - \sum_{\beta=1}^{M}\epsilon_{\beta} \sum_{j=1}^{N} \mathscr{L}^{[\beta]}_{ij} J_{2}G_{\beta}{({\bf x}_{0},{\bf x}_{0})} \delta {\bf x}_{j}  \\
			
			\;\;\;\;\;\;\; + 2 \sum_{q=1}^{L}\epsilon^{q}_{SI} k^{[q,SI]} J_{1}G^{q}_{SI}{({\bf x}_{0},{\bf x}_{0},{\bf x}_{0})} \delta {\bf x}_{i} + \sum_{q=1}^{L}\epsilon^{q}_{SI} \bigg[ \sum_{j=1}^{N} J_{2}G^{q}_{SI}{({\bf x}_{0},{\bf x}_{0},{\bf x}_{0})} \delta {\bf x}_{j} \sum_{k=1}^{N} \mathscr{A}^{[q,SI]}_{ijk} \\
			\hspace{200pt} + \sum_{k=1}^{N} J_{3} G^{q}_{SI}{({\bf x}_{0},{\bf x}_{0},{\bf x}_{0})} \delta {\bf x}_{k} \sum_{j=1}^{N} \mathscr{A}^{[q,SI]}_{ijk}\bigg].
		\end{array}
	\end{equation}
\end{widetext}
Then the symmetric property of the adjacency tensors $\mathscr{A}^{[q,SI]}$, i.e., $ \sum_{j=1}^{N} \mathscr{A}^{[q,SI]}_{ijk}=  \sum_{j=1}^{N} \mathscr{A}^{[q,SI]}_{ikj}$ and the fact that $ \sum_{k=1}^{N} \mathscr{A}^{[q,SI]}_{ijk}= K^{[q,SI]}_{ij}$, where $K^{[q,SI]}_{ij}$ counts the number of $2$-simplices to which the link $(i,j)$ participates, modifies Eq. \eqref{stability_4} as follows,
\begin{widetext}
	\begin{equation} \label{stability_5}
		\begin{array}{l}
			\delta \dot{{\bf x}}_{i}=JF({\bf x}_{0}) \delta{\bf x}_{i} + \sum_{\beta=1}^{M} \epsilon_{\beta} k^{\beta} \bigg[J_{1}G_{\beta}{({\bf x}_{0},{\bf x}_{0})} + J_{2}G_{\beta}{({\bf x}_{0},{\bf x}_{0})} \bigg] \delta {\bf x}_{i} - \sum_{\beta=1}^{M}\epsilon_{\beta} \sum_{j=1}^{N} \mathscr{L}^{[\beta]}_{ij} J_{2}G_{\beta}{({\bf x}_{0},{\bf x}_{0})} \delta {\bf x}_{j}  \\\\
			
			\;\;\;\;\;\;\;\;\;\;\;\;\; + 2\sum_{q=1}^{L} \epsilon^{q}_{SI} k^{[q,SI]} \bigg[ J_{1}G^{q}_{SI}{({\bf x}_{0},{\bf x}_{0},{\bf x}_{0})}+ J_{2}G^{q}_{SI}{({\bf x}_{0},{\bf x}_{0},{\bf x}_{0})} + J_{3}G^{q}_{SI}{({\bf x}_{0},{\bf x}_{0},{\bf x}_{0})} \bigg] \delta {\bf x}_{i}  \\\\
			
			\;\;\;\;\;\;\;\;\;\;\;\;\;  - \sum_{q=1}^{M}\epsilon^{q}_{SI} \sum_{j=1}^{N} \mathscr{L}^{[q,SI]}_{ij} \bigg[  J_{2}G^{q}_{SI}{({\bf x}_{0},{\bf x}_{0},{\bf x}_{0})} + J_{3} G^{q}_{SI}{({\bf x}_{0},{\bf x}_{0},{\bf x}_{0})}\bigg] \delta {\bf x}_{j}, 
		\end{array}
	\end{equation}
\end{widetext}
where $\mathscr{L}^{[q,SI]}$ illustrates the generalized zero row-sum Laplacian matrix associated with three-body interactions, defined by 
\begin{equation*}
	\begin{array}{l}
		\mathscr{L}^{[q,SI]}_{ij}= \begin{cases}
			-{K}^{[q,SI]}_{ij}, & i \neq j \\
			2k^{[q,SI]}_{i}, & i=j.
		\end{cases}
	\end{array}
\end{equation*}
Let us now rewrite the Eq. \eqref{stability_5} in vector form by introducing $\delta \mathbf{X}= [\delta \mathbf{x}_{1}^{tr}, \delta \mathbf{x}_{2}^{tr}, \cdots, \delta \mathbf{x}_{N}^{tr}]^{tr}$, where $[\;\;]^{tr}$ denotes the matrix transpose and denoting the terms as $h=JF({\bf x}_{0})$, $H_{\beta}=J_{1}G_{\beta}{({\bf x}_{0},{\bf x}_{0})} + J_{2}G_{\beta}{({\bf x}_{0},{\bf x}_{0})}$, $\Phi_{\beta}= J_{2}G_{\beta}{({\bf x}_{0},{\bf x}_{0})}$, $\Psi_{q}=J_{1}G^{q}_{SI}{({\bf x}_{0},{\bf x}_{0},{\bf x}_{0})}+ J_{2}G^{q}_{SI}{({\bf x}_{0},{\bf x}_{0},{\bf x}_{0})} + J_{3}G^{q}_{SI}{({\bf x}_{0},{\bf x}_{0},{\bf x}_{0})}$, $\chi_{q}= J_{2}G^{q}_{SI}{({\bf x}_{0},{\bf x}_{0},{\bf x}_{0})} + J_{3} G^{q}_{SI}{({\bf x}_{0},{\bf x}_{0},{\bf x}_{0})}$. Eventually, the equation in terms of stake vectors becomes as follows,       
\begin{widetext}
	\begin{equation} \label{stability_6}
		\begin{array}{l}
			\delta \dot{{\bf X}}= \bigg[(I_{N} \otimes h)  + \sum_{\beta=1}^{M}\epsilon_{\beta} k^{\beta} (I_{N} \otimes H_{\beta}) - \sum_{\beta=1}^{M}\epsilon_{\beta} (\mathscr{L}^{[\beta]} \otimes \Phi_{\beta})  + 2 \sum_{q=1}^{L}\epsilon^{q}_{SI} k^{[q,SI]} (I_{N} \otimes \Psi_{q})
		   \\ ~~~~~~~~~~~~~~~~~~~~~~~~~~~~~~	- \sum_{q=1}^{L}\epsilon^{q}_{SI} (\mathscr{L}^{[q,SI]} \otimes \chi_{q}) \bigg] \delta {\bf X},
		\end{array}
	\end{equation} 
\end{widetext}
where $\otimes$ and $I_{N}$ symbolize the Kronecker product and $N$-dimensional identity matrix, respectively. This linearized set of variational Eq. \eqref{stability_6} has two parts: one for motion along the synchronization manifold, termed parallel modes, and another for motion across the manifold, called transverse modes. All transverse modes must converge to zero in time for the synchronization state to be stable. The linear stability analysis is then carried out by decoupling the variational equation into parallel and transverse modes and determining whether or not the latter is extinct. Now all the Laplacian matrices $\mathscr{L}^{[\beta]}$ $(\beta=1,2,\cdots,M)$ as well $\mathscr{L}^{[q,SI]}$ $(q=1,2,\cdots,L)$ are zero row-sum real symmetric matrices. Hence they are diagonalizable, and all their eigenvalues are non-negative real numbers with one common smallest eigenvalue $\gamma_{1}=0$, and the set of eigenvectors form an orthonormal basis of $\mathbb{R}^{N}$. To separate transverse and parallel modes from Eq. \eqref{stability_6}, we project the stack variable $\delta \mathbf{X}$ onto the basis of eigenvectors $V^{[1]}=\{v_{1},v_{2},\cdots,v_{N}\}$ corresponding to $\mathscr{L}^{[1]}$, the Laplacian associated with layer-$1$ by defining new variable $\eta=(V^{[1]}\otimes I_{p})^{-1}\delta \mathbf{X}$, where $v_{1}=\big(\frac{1}{\sqrt{N}},\frac{1}{\sqrt{N}},\cdots,\frac{1}{\sqrt{N}}\big)^{tr}$ is the eigenvector corresponding to the smallest eigenvalue $\gamma_{1}=0$. The set of eigenvectors chosen is fully arbitrary; any other basis can be chosen, and all other sets of eigenvectors will ultimately change to it via unitary matrix transformation. Then in terms of the new variable, the variational Eq. \eqref{stability_6} transforms as follows, 
\begin{widetext}
	\begin{equation} \label{stability_7}
		\begin{array}{l}
			\dot{\eta}= \bigg[(I_{N} \otimes h)  + \sum_{\beta=1}^{M}\epsilon_{\beta} k^{\beta} (I_{N} \otimes H_{\beta}) - \epsilon_{1}(\Gamma\otimes\Phi_{1})-\sum_{\beta=2}^{M}\epsilon_{\beta} (\tilde{\mathscr{L}}^{[\beta]} \otimes \Phi_{\beta}) 
			\\ ~~~~~~~~~~~~~~~~  + 2 \sum_{q=1}^{L}\epsilon^{q}_{SI} k^{[q,SI]} (I_{N} \otimes \Psi_{q})- \sum_{q=1}^{L}\epsilon^{q}_{SI} (\tilde{\mathscr{L}}^{[q,SI]} \otimes \chi_{q}) \bigg] \eta,
		\end{array}
	\end{equation} 
\end{widetext}
where ${V^{[1]}}^{-1}\mathscr{L}^{[1]}V^{[1]}=\Gamma=diag\{\gamma_{1}=0,\gamma_{2},\cdots,\gamma_{N}\}$, ${V^{[1]}}^{-1}\mathscr{L}^{[\beta]}V^{[1]}= \tilde{\mathscr{L}}^{[\beta]}$ and $\tilde{\mathscr{L}}^{[q,SI]}={V^{[1]}}^{-1}\mathscr{L}^{[q,SI]}V^{[1]}$.
Now, as all the Laplacians are zero row-sum matrices, the transformed Eq. \eqref{stability_7} can be decoupled
into two parts as follows
\begin{widetext}
	\begin{subequations} 
		\begin{eqnarray}
			\dot{\eta}_{1}= (h+\sum_{\beta=1}^{M}\epsilon_{\beta}k^{\beta}H_{\beta}+2\sum_{q=1}^{L}\epsilon^{q}_{SI}k^{[q,SI]}\Psi_{q})\eta_{1}, \label{stability_8a}\\
			\dot{\eta}_{i}=(h+\sum_{\beta=1}^{M}\epsilon_{\beta}k^{\beta}H_{\beta}+2\sum_{q=1}^{L}\epsilon^{q}_{SI}k^{[q,SI]}\Psi_{q}-\epsilon_{1}\gamma_{i}\Phi_{1})\eta_{i}-\sum_{\beta=2}^{M}\epsilon_{\beta}\sum_{j=2}^{N}\tilde{\mathscr{L}}^{[\beta]}_{ij}\Phi_{\beta}\eta_{j}- \sum_{q=1}^{L}\epsilon^{q}_{SI}\sum_{j=2}^{N}\tilde{\mathscr{L}}^{[q,SI]}_{ij}\chi_{q}\eta_{j} \label{stability_8b},
		\end{eqnarray}
	\end{subequations}
\end{widetext}
where $\eta_{1}$ indicates the parallel mode and the transverse modes are symbolized by $\eta_{i}$, $i=2,3,\cdots,N$. 
Therefore, the problem of synchronization stability is reduced in solving the coupled linear differential Eq. \eqref{stability_8b} associated with transverse mode to calculate the maximum Lyapunov exponent (MLE). The Eq. \eqref{stability_8b} is called the master stability equation. For the synchronization state to be stable, the MLE must be negative.     
\par Due to more complexity of multiple layers and higher-order structures, the transverse variational Eq. \eqref{stability_8b} is a $(N-1)p$-dimensional linear coupled differential equation, and in general, it is difficult to decouple it to lower dimensional form. However, there is a suitable instance for which the coupled transverse modes can be optimally separated, and as a result, the $(N-1)p$-dimensional coupled equation transforms into $(N-1)$ numbers of $p$-dimensional linear differential equations. The relevant instance is as follows-
\par {\it When all the pairwise Laplacians $\mathscr{L}^{[\beta]}$ and the generalized Laplacians $\mathscr{L}^{[q,SI]}$ commute with each other then the transverse Eq. \eqref{stability_8b} can be decoupled into $(N-1)$ numbers of $p$-dimensional equations} as,
\begin{equation} \label{stability_9}
	\begin{array}{l}
		\dot{\eta}_{i}= \bigg(h+\sum_{\beta=1}^{M}\epsilon_{\beta}k^{\beta}H_{\beta}+2 \sum_{q=1}^{L}\epsilon^{q}_{SI}k^{[q,SI]}\Psi_{q}
		\\\\  -\sum_{\beta=1}^{M}\epsilon_{\beta}\gamma^{\beta}_{i}\Phi_{\beta}- \sum_{q=1}^{L}\epsilon^{q}_{SI}\gamma_{i}^{[q,SI]}\chi_{q}\bigg)\eta_{i},~ i=2,3,\cdots,N,
	\end{array}
\end{equation}
where ${V^{[1]}}^{-1}\mathscr{L}^{[\beta]}V^{[1]}=\Gamma^{\beta}=diag\{0=\gamma^{\beta}_{1}<\gamma^{\beta}_{2}\cdots\leq \gamma^{\beta}_{N}\}$ and ${V^{[1]}}^{-1}\mathscr{L}^{[q,SI]}V^{[1]}=\Gamma^{[q,SI]}=diag\{0=\gamma^{[q,SI]}_{1}<\gamma^{[q,SI]}_{2}\cdots\leq \gamma^{[q,SI]}_{N}\}$, as in this scenario each of the Laplacians are diagonalizable with respect to the eigenvector matrix of the Laplacian of 1st layer $\mathscr{L}^{[1]}$.
\section{Numerical validation} \label{numericals}
A series of outcomes is presented below, all of which support the validity and broad applicability of our method. Here we consider two three-dimensional chaotic oscillators, namely the Sherman neuronal model \cite{belykh_sherman,sherman1994} associated with pancreatic $\beta$-cells and the R\"{o}ssler oscillators \cite{rossler1976equation}, coupled through various nonlinear pairwise and non-pairwise coupling functions interacting via two $(M=2)$ distinct connection layers. For numerical simulations, we investigate the complete synchronization of the simplicial complex composed of $N$ nodes. We employ the synchronization error to analyze complete synchronization as, 
\begin{equation} \label{error}
	\begin{array}{l}
		E=\lim\limits_{T\to\infty}\dfrac{1}{T}\bigintss_{t_{trans}}^{t_{trans}+T} \sum\limits_{i,j=1}^{N} \dfrac{\|\mathbf{x}_{j}-\mathbf{x}_{i}\|}{N(N-1)},
	\end{array} 
\end{equation}  
where  $\|\cdot\|$ symbolizes the Euclidean norm, $t_{trans}$ determines the transient of the numerical simulation, and $T$ is a sufficiently large positive number. The zero value of the synchronization error $E$ indicates the emergence of complete synchrony. The simplicial complex \eqref{gen_model} is integrated using the $4$th order Runge-Kutta scheme subject to integration time step $dt=0.001$ for neuron dynamics and $dt=0.01$ for R\"{o}ssler system, over a period of $t=50000$ time units with a transient of $40000$ time units. Each dynamical node's initial condition is randomly selected from the solitary node dynamics phase space.   
\par Throughout this section, we study the impact of higher-order interactions on complete synchrony with arbitrary coupling mechanisms and verify all the results with our analytical conjectures. In a chaotic bursting regime, we select the system parameters of individual nodes and focus on variation of coupling strengths along with different coupling schemes and network topologies.
\subsection{Application to neuron dynamics}
Recent studies in neuroscience have revealed that neurons may communicate through many-body interactions \cite{beyond_pairwise,parastesh2022,ince2009presence,amari2003synchronous}. Specifically, astrocytes and other glial cells are considered a possible source of
many-body interactions \cite{fellin2004neuronal,tlaie2019high}, as they make contact with thousands of synapses and actively modulate their functions \cite{allen2009glia}. Although it is still unclear how to account for these many-body interactions in nonlinear neuronal models, we think our simplicial complex framework may suitably fit the neuronal network systems to study synchronization in the presence of many-body interactions. 
\par One neuron generally transfers information to other neurons through gap junctions and chemical synapses. A signal is sent chemically through a chemical synapse by neurochemical compounds like acetylcholine, gamma-aminobutyric acid, dopamine, and serotonin, which are contained inside microscopic synaptic vesicles. Exocytoses release neurotransmitters probabilistically from a presynaptic neuron into the synaptic gap. Following that, these compounds attach to certain receptor molecules in nearby postsynaptic neuronal cells. The gap between the pre-and postsynaptic ends could be as much as 20–40 nm \cite{hormuzdi2004electrical} in this scenario. A gap junction connects the cytoplasm of neighboring cells in electrical synapses. Electric current, calcium, cyclic AMP, and inositol-1,4,5 triphosphate pass in both directions between the presynaptic end and the postsynaptic neuron. The pre-and postsynaptic neurons' membranes are extraordinarily close to each other, about 3.5 nm \cite{kandel2000principles}, and signal transmission is significantly faster than chemical transmission. In most neurological systems, both types of synapses exist. In interneuronal transmission, both types of interaction may not always be present at all times; rather, they operate independently \cite{pereda2014electrical} throughout time. A mixed synapse occurs when two neurons communicate through both chemical and electrical synapses. A heterosynaptic connection, on the other hand, occurs when a neuron is coupled to two different neurons, one by a chemical synapse and the other through an electrical synapse.   			
\par Here we analyze a group of pancreatic $\beta$ cells, represented by
the paradigmatic Sherman model \cite{belykh_sherman,sherman1994}, subjected to both pairwise and three-body interactions. The velocity profile of a single $\beta$ cell is given by 
\begin{equation*}
	\begin{array}{l}
		F(\mathbf{x})= F \begin{bmatrix}
			V \\ n\\ s
		\end{bmatrix} \\
		=\begin{bmatrix}
			f(V,n,s)=\frac{1}{\tau}\{-I_{Ca}(V)-I_{K}(V,n)-I_{S}(V,s)\} \\
			g(V,n,s) =\frac{1}{\tau} \{\mu [n^{\infty}(V)-n]\} \\
			h(V,n,s) =\frac{1}{\tau_s} \{s^{\infty}(V)-s\}
		\end{bmatrix},
	\end{array}
\end{equation*}
where two fast currents: calcium $I_{Ca}$,  persistent potassium $I_{K}$, and a slow potassium current $I_{S}$ are given by $I_{S}(V,s)=g_{S}s(V-E_{K})$, $I_{K}(V,n)=g_{K}n(V-E_{K})$, and $I_{Ca}(V)=g_{Ca}m^{\infty}(V-E_{Ca})$, respectively.
$V$ is the membrane potential corresponding to the reversal potential $E_{K}=-0.075$, $E_{Ca}=0.025$. $m$, $n$, and $s$ are the voltage-dependent gating variables. The maximum conductance and time constants are $g_{Ca}=3.6$, $g_{K}=10$, $g_{S}=4$ and $\tau=0.02$, $\tau_{s}=5$. $\mu=1$, an auxiliary scaling factor, manages the time scale of the persistent potassium channels. The values of the gating variables at steady state are 
\begin{equation*}
	\begin{array}{l}
		m^{\infty}(V)=\{1+\exp[-83.34(V+0.02)]\}^{-1}, \\n^{\infty}(V)=\{1+\exp[-178.57(V+0.016)]\}^{-1}, \;\mbox{and}\\ s^{\infty}(V)=\{1+\exp[-100(V+0.035345)]\}^{-1}.
	\end{array}	
\end{equation*}
The parameter values are taken in such a way that the isolated node dynamics exhibits chaotic behavior (spiking bursting). 
\par We assume that the neurons are interconnected concurrently through two (M = 2) distinct interacting layers, subject to gap junction coupling via electrical synapses and chemical ion transit via chemical synapses. Therefore, the pairwise connections in the layers are described by the coupling functions $G_1(\mathbf{x}_{i},\mathbf{x}_{j})=[V_{j}-V_{i}, 0, 0]^{tr}$ to describe gap junctions, and $G_2(\mathbf{x}_{i},\mathbf{x}_{j})=[(E_{S}-V_{i})\Gamma(V_{j}), 0, 0]^{tr}$  to describe chemical ion transportation through chemical synapses. For three-body interactions, we consider two different instances: (1) only the layer with electrical synapse coupling allows three-body connection, described by the diffusive coupling function $G^{e}_{SI}(\mathbf{x}_{i},\mathbf{x}_{j},\mathbf{x}_{k})=[V_{j}+V_{k}-2V_{i}, 0, 0]^{tr}$ and (2) three-body interactions are permitted only in the layer with chemical synapse coupling, represented by the nonlinear coupling form $G^{c}_{SI}(\mathbf{x}_{i},\mathbf{x}_{j},\mathbf{x}_{k})=[(E_{S}-V_{i})\{\Gamma(V_{j})+\Gamma(V_{k})\}, 0, 0]^{tr}$. Here, the sigmoidal input-output function is, 
\begin{eqnarray*}
	\Gamma(V)=\frac{1}{1+\exp(\lambda_{s}(V-\Theta_s))},
\end{eqnarray*} describes the procedure for activation and deactivation of nonlinear chemical synapse. The synaptic reversal potential is $E_s=-0.02$. Slope of the sigmoidal function, and synaptic firing threshold are determined by the real-valued constants $\lambda_{s}$ and $\Theta_{s}$, which are fixed at $\Theta_s=-0.045$ and $\lambda_{s}=-1000$. Therefore, the equation of motions describing the dynamics of the neuronal simplicial complex for two instances are given by,
  	\begin{equation} \label{hoi_elec}
	 \begin{array}{l}
	\tau\dot{V}_{i}=-I_{Ca}(V_{i})-I_{K}(V_{i},n_{i})-I_{S}(V_{i},s_{i})\\
		~~~~~~ +\epsilon \sum\limits_{j=1}^{N}{\mathscr{A}_{ij}^{[e]}}(V_{j}-V_{i}) +g_{c}(E_{S}-V_{i})\sum\limits_{j=1}^{N}{\mathscr{A}_{ij}^{[c]}}\Gamma(V_{j}) \\
		 ~~~~~~~~~~~~~~~~~~~ +\epsilon^{e}_{SI}\sum\limits_{j=1}^{N}\sum\limits_{k=1}^{N}{\mathscr{A}_{ijk}^{[e,SI]}}(V_{j}+V_{k}-2V_{i}), \\
		\tau\dot{n}_{i}=\mu [n^{\infty}(V_{i})-n_{i}], \\
		\tau_{s}\dot{s}_{i}=s^{\infty}(V_{i})-s_{i}, 
	\end{array}
\end{equation} 		  	
and \begin{equation} \label{hoi_chem}
	\begin{array}{l}
		\tau\dot{V}_{i}=-I_{Ca}(V_{i})-I_{K}(V_{i},n_{i})-I_{S}(V_{i},s_{i}) \\
		~~~~~~+\epsilon \sum\limits_{j=1}^{N}{\mathscr{A}_{ij}^{[e]}}(V_{j}-V_{i}) +g_{c}(E_{S}-V_{i})\sum\limits_{j=1}^{N}{\mathscr{A}_{ij}^{[c]}}\Gamma(V_{j}) \\
		~~~~~~~~~~~ +\epsilon^{c}_{SI}(E_{S}-V_{i})\sum\limits_{j=1}^{N}\sum\limits_{k=1}^{N}{\mathscr{A}_{ijk}^{[c,SI]}}[\Gamma({V}_{j})+\Gamma({V}_{k})], \\
		\tau\dot{n}_{i}=\mu [n^{\infty}(V_{i})-n_{i}], \\
		\tau_{s}\dot{s}_{i}=s^{\infty}(V_{i})-s_{i},
	\end{array}
    \end{equation}
where $\epsilon$, $g_{c}$ symbolize electrical and chemical synaptic coupling strengths, and $\epsilon^{e}_{SI}$, $\epsilon^{c}_{SI}$ are the corresponding higher-order coupling strengths. The adjacency matrices $\mathscr{A}^{[e]}$ and $\mathscr{A}^{[c]}$ describes the pairwise network connectivity of electrical and chemical synapses, and $\mathscr{A}^{[e,SI]}$, $\mathscr{A}^{[c,SI]}$ represent adjacency tensors corresponding to three-body interactions. In general, the connectivity network of electrical and chemical synapses are taken to be bidirectional and unidirectional, respectively. But for our case, we consider both the connectivity network to be bidirectional. Furthermore, since the coupling forms associated with chemical ion transportation are nonlinear synchronization invasive (i.e., $G_{2}(\mathbf{x},\mathbf{x}) \neq 0$ and $G^{c}_{SI}(\mathbf{x},\mathbf{x},\mathbf{x}) \neq 0$), we consider identical node degree $k_{i}^{c}=k^{c}$ corresponding to $1$-simplex and $k_{i}^{[c,SI]}=k^{[c,SI]}$, the number of $2$-simplices adjacent to a node for synchronization invariance condition to be satisfied. However, since the coupling forms corresponding to gap junctions are linear diffusive, no further restriction on connection topology is needed in this scenario.   
	\begin{figure*}[ht] 
	\centerline{
		\includegraphics[scale=0.3]{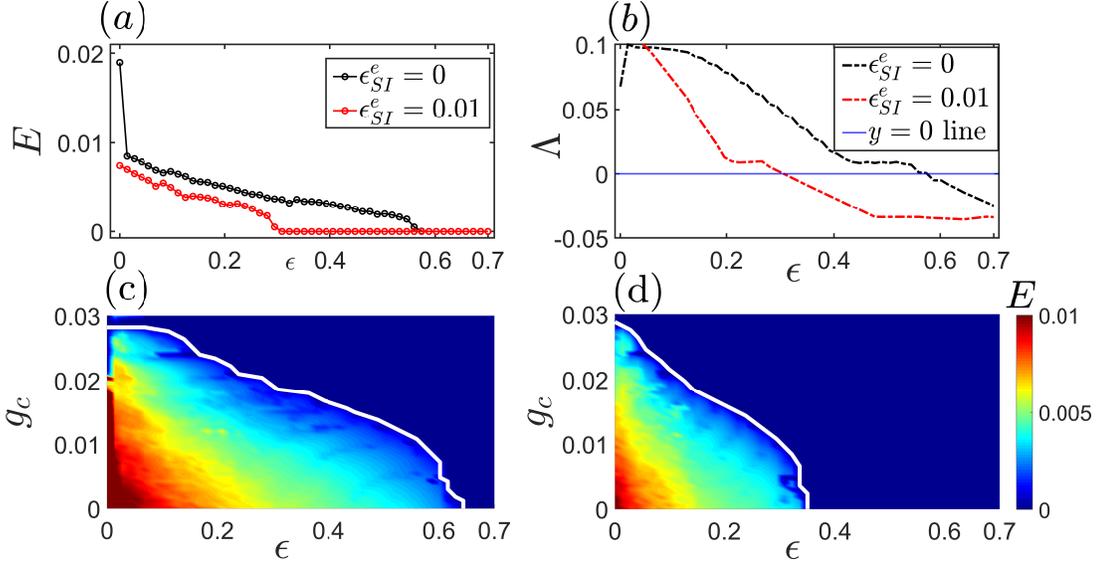}}
	\caption{{\bf Synchronization in coupled Sherman model with higher-order gap junction coupling.} (a) Synchronization error $E$ as a function of pairwise gap junction coupling strength $\epsilon$ for a fixed value of pairwise chemical synaptic coupling $g_{c}=0.01$. The synchronization error in the absent $(\epsilon^{e}_{SI}=0)$ and present $(\epsilon^{e}_{SI}=0.01)$ of higher-order gap junction coupling is displayed by black and red circle curves, respectively. The dashed black and red lines in (b) depict the corresponding maximum Lyapunov exponent curves. The variation of synchronization error $E$ in $(\epsilon, g_{c})$ parameter planes for (c) $\epsilon^{e}_{SI}=0$ and (d) $\epsilon^{e}_{SI}=0.01$ are represented. The solid white curves in (c) and (d) denote the theoretically derived synchronization threshold curves obtained from Eq. \eqref{elec_trans}. Here the other parameter values are $k_{c}=4$, $k_{sw}=4$ and $p_{sw}=0.15$.}
	\label{sw_random}
\end{figure*}
\subsubsection{Higher-order gap junction coupling}
We first consider the layer with a connection through electrical synapses to allow three-body interactions among the neurons. Therefore, the equations of motion governing the dynamics of our simplicial complex satisfy Eq. \eqref{hoi_elec}. We consider the connectivity topology of electrical synapses to be small-world \cite{muldoon2016small}, given by the adjacency matrix $\mathscr{A}^{[e]}$. Particularly, we use the Watts–Strogatz (WS) graph model \cite{watts1998collective}, which starts with a non-local ring with $N=200$ nodes, $k_{sw}=4$ nearest neighbor on each side, and the edge rewiring probability $p_{sw}=0.15$. The $2$-simplices are constructed by promoting all the empty triangles of the small-world network to full triangles, represented by the adjacency tensor $\mathscr{A}^{[SI]}$. On the other hand, the chemical synaptic network is considered to be a random network, described by the adjacency matrix $\mathscr{A}^{[c]}$ with constant node degree $k_{c}=4$. 
\par We begin by examining the development of synchronization error $E$ (given by \eqref{error}) for varying coupling strengths to understand the emergence of neuronal synchrony in our simplicial complex \eqref{hoi_elec}. Figure \ref{sw_random}(a) shows the variation in $E$ as a function of the pairwise electrical synaptic coupling strength $\epsilon$ for various values of higher-order coupling strength $\epsilon^{e}_{SI}$, with fixed chemical synaptic coupling $g_{c}=0.01$. The figure delineates two different curves corresponding to two different values of $\epsilon^{e}_{SI}$. Solid black circle corresponds synchronization error for $\epsilon^{e}_{SI}=0$, i.e., when the higher-order interactions are not considered. The critical point to achieve synchronization is $\epsilon \approx 0.58$ in this scenario. While for a non-zero value of higher-order coupling $\epsilon^{e}_{SI}=0.01$ ( shown in the red circles ), the synchrony is achieved at a much lower value of the coupling $\epsilon=0.35$. {\it Therefore, our observation indicates that synchronization can be improved in the neuronal simplicial complex by introducing three-body interactions through electrical synaptic coupling.} 
\par To verify our numerical result, we then proceed through the linear stability analysis of the synchronization state based on the MSF approach. Followed by a series of calculations detailed in Sec. \ref{stability}, the variational equation transverse to synchronization manifold can be written analogous to Eq. \eqref{stability_8b} as,  
\begin{equation} \label{elec_trans}
	\begin{array}{l}
		\tau\delta\dot{V}_{i}=  Jf(\mathbf{x}_{0})\delta \mathbf{x}_{i}-\epsilon \gamma_{i}\delta V_{i}- 2\epsilon^{e}_{SI}\sum\limits_{j=2}^{N} \tilde{\mathscr{L}}_{ij}^{[e,SI]}\delta V_{j} \\ ~~~~~
		+g_{c} k_{c} [- \Gamma(V_{0})+(E_{S}-V_{0})\Gamma'(V_{0})] \delta V_{i} 
		\\ ~~~~~~~~~~~~  - g_{c}(E_{S}-V_{0})\sum\limits_{j=2}^{N}{\tilde{\mathscr{L}}_{ij}^{[c]}}\Gamma'(V_{0}) \delta V_{i},\\
		
		\tau\delta \dot{n}_{i}=Jg(\mathbf{x}_{0}) \delta \mathbf{x}_{i}, \\
		\tau_{s} \delta \dot{s}_{i}=Jh(\mathbf{x}_{0}) \delta \mathbf{x}_{i}, ~~~~~ i=2,3,\cdots,N,
	\end{array}
\end{equation}
where $J$ denotes the Jacobian operator, $'$ represents derivative with respect to $V$, $\gamma_{1}(=0) < \gamma_{2} \leq \cdots \leq \gamma_{N}$ are the eigenvalues of the Laplacian matrix $\mathscr{L}^{[e]}$. $\tilde{\mathscr{L}}^{[c]}$ and $\tilde{\mathscr{L}}^{[e,SI]}$ are the $(N-1)\times(N-1)$ real matrices obtained through transformation the Laplacians $\mathscr{L}^{[c]}$ and $\mathscr{L}^{[SI]}$ by the matrix of orthonormal eigenvectors of $\mathscr{L}^{[e]}$. Here $\mathbf{x}_{0}=(V_{0},n_{0},s_{0})^{tr}$ represents the states of the synchronized solution given by,
\begin{equation}\label{sync_sol1}
	\begin{array}{l}
			\tau\dot{V}_{0}=-I_{Ca}(V_{0})-I_{K}(V_{0},n_{0})-I_{S}(V_{0},s_{0}) \\~~~~~~~~~~~~~~~~~~~ +g_{c}k_{c}(E_{S}-V_{0})\Gamma(V_{0}),\\
		\tau\dot{n}_{0}=\mu [n^{\infty}(V_{0})-n_{0}], \\
		\tau_{s}\dot{s}_{0}=s^{\infty}(V_{0})-s_{0}.
	\end{array}
\end{equation}  
Solving $3(N-1)$-dimensional linearized Eq. \eqref{elec_trans} along with the $3$-dimensional nonlinear Eq. \eqref{sync_sol1} for the calculation of maximum Lyapunov exponent gives the condition for stable synchronization state. Let $\Lambda$ be the maximum Lyapunov exponent transverse to the synchronization manifold. Then the stable synchronous solution is obtained if $\Lambda$ becomes negative with varying coupling strengths. Figure \ref{sw_random}(b) portrays the variation of $\Lambda$ with respect to $\epsilon$ for same set of other coupling strengths $g_{c}=0.01$ and $\epsilon^{e}_{SI}=0,0.01$ used in Fig. \ref{sw_random}(a). The dashed black and red curves display the variation of $\Lambda$ for $\epsilon^{e}_{SI}=0$ and $\epsilon^{e}_{SI}=0.01$, respectively. It is clearly observable that both the curves cross the zero line and become negative exactly at the same critical values of $\epsilon$ where the corresponding curves of synchronization error $E$ achieved zero value as reported in Fig. \ref{sw_random}(a). {\it Hence it affirms our obtained result about the enhancement of synchrony in the neuronal simplicial complex by the inclusion of higher-order electrical synaptic coupling.}
\par Thereafter, we concentrate on the effect of higher-order interaction in the $(\epsilon,g_{c})$ parameter plane by varying pairwise electrical and chemical synaptic coupling strengths for fixed values of non-pairwise electrical synaptic coupling parameter $\epsilon^{e}_{SI}$. We plot the corresponding synchronization error $E (\epsilon,g_{c})$ in Figs. \ref{sw_random}(c) and \ref{sw_random}(d), where colorbar indicates the variation of $E$. Figure \ref{sw_random}(c) depicts how synchronization emerges for concurrent variation in $\epsilon \in [0,0.7]$ and $g_{c} \in [0,0.03]$ whenever there is no higher-order coupling, i.e., $\epsilon^{e}_{SI}=0$. As we can notice that as $g_{c}$ is increased, lower values of $\epsilon$ are found to be sufficient for the achievement of complete synchrony. A similar phenomenon is observed when we increase $\epsilon^{e}_{SI}$ to $0.01$, depicted in Fig. \ref{sw_random}(d). But in this case, the critical values of $\epsilon$ for increasing $g_{c}$ are sufficiently smaller than in the previous scenario. The threshold for synchrony lowers down from $\epsilon \approx 0.66$ to $\epsilon \approx 0.35$ even with no chemical synaptic effect, i.e., when $g_{c}=0$. However, the threshold value of $g_{c}$ remains the same in the absence of pairwise electrical synaptic coupling, i.e., $\epsilon=0$. The solid white curves in Figs. \ref{sw_random}(c) and \ref{sw_random}(d) are theoretical predictions of the synchronization thresholds by means of maximum Lyapunov exponents obtained from Eq. \eqref{elec_trans}, which indeed fully matches with the numerical results obtained by solving Eq. \eqref{hoi_elec} for synchronization error $E$.
 \begin{figure*}[ht] 
	\centerline{
		\includegraphics[scale=0.05]{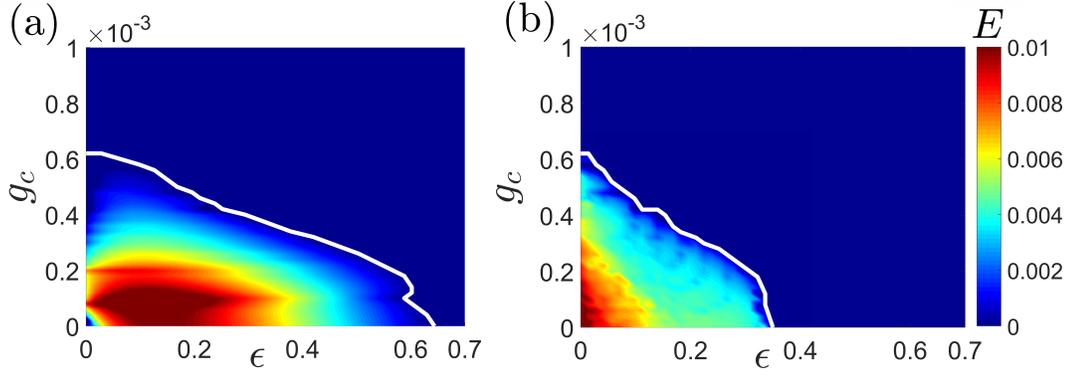}}
	\caption{{\bf Synchronization in coupled Sherman model with higher-order gap junction coupling and all-to-all pairwise chemical synaptic connections.} Synchronization error $E$ in $(\epsilon, g_{c})$ plane for (a) $\epsilon^{e}_{SI}=0$, and (b) $\epsilon^{e}_{SI}=0.01$. The other parameter values are $p_{sw}=0.15$ and $k_{sw}=4$.  The solid white curves depict the critical curves for which $\Lambda = 0$ obtained from Eq. \eqref{elec_trans}, while the regions below and above it denote the unstable and stable synchronization states, respectively.}
	\label{sw_all_1}
\end{figure*}
\par Now we investigate the effect of higher-order electrical synaptic coupling on neuronal synchronization by changing the network topology of the layer interacting via chemical synapses. Therefore, we consider that each neuron is connected to all the other neurons through chemical synapses, i.e., the connection topology of the layer interacting through chemical synapses is considered to be an all-to-all network. Figure \ref{sw_all_1} represents the results for variation of synchronization error $E$ in $(\epsilon,g_{c})$ parameter plane for two values of higher-order coupling strength- $\epsilon^{e}_{SI}=0$, i.e., when no higher-order effect is considered (Fig. \ref{sw_all_1}(a)) and a non-zero value of higher-order electrical synaptic coupling $\epsilon^{e}_{SI}=0.01$ (Fig. \ref{sw_all_1}(b)). It is clearly observable that the phenomena remain almost the same as discovered previously in Fig. \ref{sw_random}, i.e., the critical value of $\epsilon$ for synchrony decreases with increasing $g_{c}$. Further, the critical values for $\epsilon$ remain almost the same, only the critical values of $g_{c}$ decrease as the number of chemical synaptic connections increases in this scenario. 

\begin{figure}[ht] 
	\centerline{
		\includegraphics[scale=0.35]{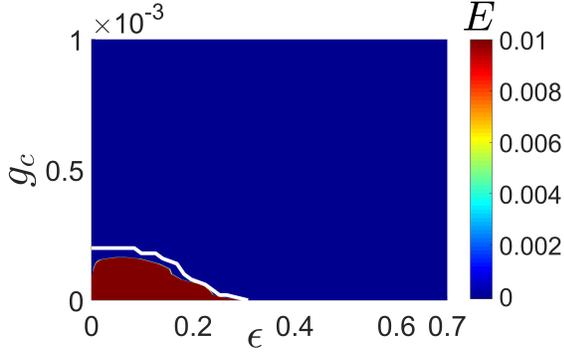}}
	\caption{{ \bf Synchronization in coupled Sherman model with higher-order chemical synaptic coupling.} Synchronization error $E$ in $(\epsilon, g_{c})$ parameter plane for fixed value of higher-order chemical synaptic coupling strength $\epsilon^{c}_{SI}=10^{-6}$. The other parameter values are $k_{sw}=4$ and $p_{sw}=0.15$.  The solid white curve depicts the critical curve for which $\Lambda= 0$ obtained from Eq. \eqref{ehem_trans}, while the regions below and above it denote the unstable and stable synchronization states, respectively.}
	\label{sw_all_2}
\end{figure}
\subsubsection{Higher-order chemical synaptic coupling}    
We now consider that the layer interacting through chemical synapses allows the three-body interactions among neurons, and the layer interacting via gap junction allows only pairwise connections. Then the equation of motion governing the dynamics of the simplicial complex is given by Eq. \eqref{hoi_chem}. Now one can notice that both the pairwise and non-pairwise coupling functions corresponding to chemical ion transportation are nonlinear in form and further $G_{2}(\mathbf{x},\mathbf{x}) \neq 0$, $G^{c}_{SI}(\mathbf{x},\mathbf{x},\mathbf{x}) \neq 0$. So to achieve complete neuronal synchrony according to the invariance condition \eqref{invariance_2}, the topology of the layer interacting through chemical synapses should qualify further restriction about the same node degree $k_{i}^{c}=k^{c}$ and $k_{i}^{[c,SI]}=k^{[c,SI]}$. We, therefore, consider the simplest configuration that the neurons are all-to-all connected. We consider this scenario so that we can compare the effect of higher-order chemical synaptic coupling with the higher-order electrical synaptic coupling discussed in the previous section (Fig. \ref{sw_all_1}). One can choose other network topologies that satisfy the synchronization invariance condition, for example, the non-local network also fulfills the demand of the invariance condition.
\par Figure \ref{sw_all_2} delineates the synchrony and desynchrony regions in terms of synchronization error $E$ in $(\epsilon,g_{c})$ parameter plane for fixed non-pairwise chemical synaptic coupling strength $\epsilon^{c}_{SI}=10^{-6}$. It is easily discoverable that the synchronization region is greatly enhanced in this case. Furthermore, in the presence of higher-order chemical synaptic coupling, not only the critical value of $g_{c}$ lowers down when $\epsilon=0$ but also the critical value of $\epsilon$ decreases when $g_{c}=0$, which is not the case with only higher-order electrical synaptic coupling as elucidated in Fig. \ref{sw_all_1}(b). In that scenario, critical value of $g_{c}$ when $\epsilon=0$ is unaffected with increasing non-pair electrical synaptic coupling strength $\epsilon^{e}_{SI}$ (Figs. \ref{sw_all_1}(a), \ref{sw_all_1}(b)). 
\par Now, since the neurons are connected globally through the chemical synapses, the pairwise and generalized Laplacians $\mathscr{L}^{[c]}$ and $\mathscr{L}^{[c,SI]}$ must satisfy the relation, $\mathscr{L}^{[c,SI]}= (N-2) \mathscr{L}^{[c]}$. Furthermore, the Laplacians $\mathscr{L}^{[c]}$ and $\mathscr{L}^{[e]}$ commute with each other. So the master stability Eq. \eqref{stability_8b} decouples optimally in this instance, and as a result, the transverse variational equation can be written analogous to Eq. \eqref{stability_9} as,
\begin{equation} \label{ehem_trans}
	\begin{array}{l}
		\tau\delta\dot{V}_{i}=  Jf(\mathbf{x}_{0})\delta \mathbf{x}_{i}+ \bigg\{g_{c}  (N-1)[-\Gamma(V_{0})+\Gamma'(V_{0})(E_{S}-V_{0})]\\\\ +2(N-1)(N-2)\epsilon^{c}_{SI}[-\Gamma(V_{0})+\Gamma'(V_{0})(E_{S}-V_{0})]-\epsilon \gamma^{e}_{i}\\\\
		-g_{c} \gamma^{c}_{i}\Gamma'(V_{0})(E_{S}-V_{0})-2\epsilon^{c}_{SI}\gamma^{[c,SI]}_{i}\Gamma'(V_{0})(E_{S}-V_{0}) \bigg\}\delta V_{i}, \\\\
		\tau\delta \dot{n}_{i}=Jg(\mathbf{x}_{0}) \delta \mathbf{x}_{i}, \\\\
		\tau_{s} \delta \dot{s}_{i}=Jh(\mathbf{x}_{0}) \delta \mathbf{x}_{i}, ~~~~~ i=2,3,\cdots,N,
	\end{array}
\end{equation}         
where the eigenvalues of the Laplacian $\mathscr{L}^{[e]}$ are symbolized by $\gamma^{e}_{1}(=0) < \gamma^{e}_{2} \leq \cdots \leq \gamma^{e}_{N}$. $\gamma^{c}_{i}=N$ and $\gamma^{[c,SI]}_{i}=N(N-2)$, $i=2,3,\cdots,N$ are the eigenvalues of the Laplacians $\mathscr{L}^{[c]}$ and $\mathscr{L}^{[c,SI]}$, respectively. In Fig. \ref{sw_all_2}, the solid white line depicts the theoretical conjecture obtained from Eq. \eqref{ehem_trans} solving for maximum Lyapunov exponents, which confirms that the numerical result obtained by solving Eq. \eqref{hoi_chem} is in good agreement with our analytical derivation.
\begin{figure*}[ht] 
	\centerline{
		\includegraphics[scale=0.3]{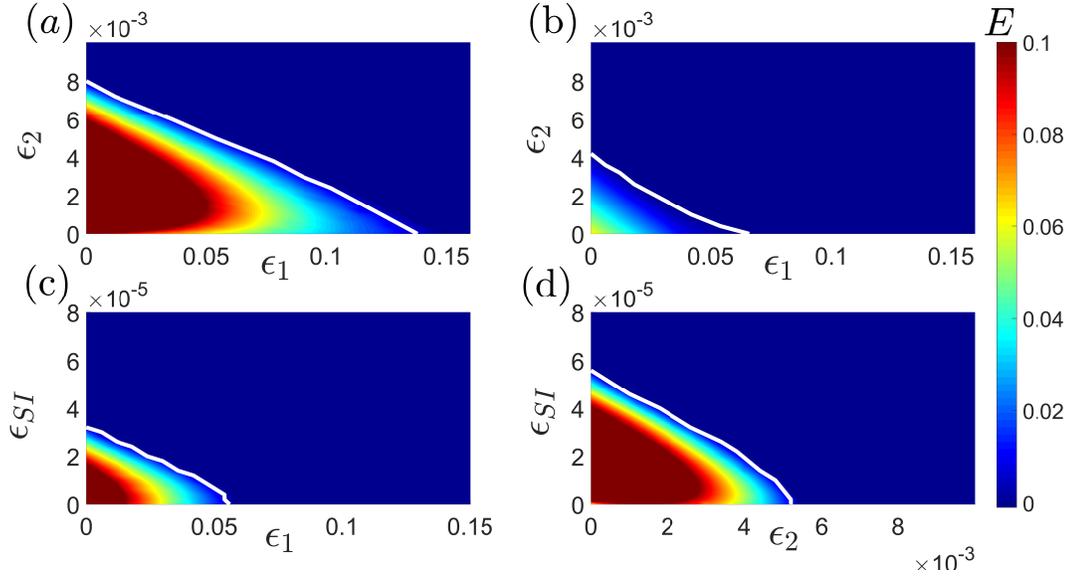}}
	\caption{{\bf Synchronization in coupled R\"{o}ssler oscillators with higher-order nonlinear coupling function.} Synchronization error $E$ in $(\epsilon_{1},\epsilon_{2})$ parameter plane for (a) $\epsilon_{SI}=0$ and (b) $\epsilon_{SI}=0.00005$. The contour plots of synchronization error $E$ in the parameter space (c) $(\epsilon_{1},\epsilon_{SI})$ and (d) $(\epsilon_{2},\epsilon_{SI})$ for fixed values of $\epsilon_{2}=0.005$ and $\epsilon_{1}=0.05$ are depicted, respectively. The solid white curves correspond to the theoretical predictions obtained from Eq. \eqref{ross_trans}.}
	\label{rossler}
\end{figure*} 
\subsection{Application to simplicial complexes of coupled R\"{o}ssler oscillators}
To further illustrate how higher-order interactions play a crucial role in the emergence of the synchronization state, we investigate the simplicial complex Eq. \eqref{gen_model} composed of $N=100$ nodes having individual node dynamics as R\"{o}ssler oscillators, interacting with nonlinear coupling functions through two $(M=2)$ distinct interaction layers, where three-body interactions are allowed in only one layer. Then the equation of motion describing the dynamics of the simplicial complex can be written as follows, 
\begin{equation}
	\begin{array}{l}
		\dot{x}_{i}= -y_{i} -z_{i}+ \epsilon_{1} \sum_{j=1}^{N} \mathscr{A}^{[1]}_{ij} (\alpha-x_{i})(x_{j}-\beta)^{2}, \\\\
        \dot{y}_{i}= x_{i} + a y_{i}+ \epsilon_{2} \sum_{j=1}^{N} \mathscr{A}^{[2]}_{ij} (\alpha-y_{i})(y_{j}-\beta)^{2}	\\\\ 
        + \epsilon_{SI} \sum_{j=1}^{N}\sum_{k=1}^{N} \mathscr{A}^{[SI]}_{ijk} (\alpha-y_{i})[(y_{j}-\beta)^{2}+(y_{k}-\beta)^{2}], \\\\
        \dot{z}_{i}= b + z_{i} (x_{i}-c), ~~~~~~~ i=1,2,\cdots,N,        
\end{array}
\end{equation}   
where $G_{1}=[(\alpha-x_{i})(x_{j}-\beta)^{2}, 0, 0]^{tr}$, $G_{2}=[0, (\alpha-y_{i})(y_{j}-\beta)^{2}, 0]^{tr}$ are the pairwise coupling forms in the layers with corresponding coupling strengths $\epsilon_{1}$, $\epsilon_{2}$ and $G_{SI}= [0, (\alpha-y_{i})\{(y_{j}-\beta)^{2}+(y_{k}-\beta)^{2}\}, 0]^{tr}$ is the functional form of the non-pairwise coupling acting in the second layer with coupling strength $\epsilon_{SI}$. The connection topology of the first layer is considered to be the random network with constant node degree $k_{1}=5$ to fulfill the synchronization invariance condition, and for the second layer, we consider the nodes are all-to-all connected. Here, the system parameters are chosen as $a=b=0.2$, $c=5.7$ for the isolated node dynamics to be in a chaotic state and the coupling parameters are taken as $\alpha=0.37$, $\beta=-0.37$.

\par Figure \ref{rossler} portrays the results corresponding to the complete synchronization by varying the coupling strengths in two-dimensional parameter planes in terms of synchronization error $E$. In the absence of higher-order coupling $(\epsilon_{SI}=0)$, the synchronization and de-synchronization region is depicted in $(\epsilon_{1},\epsilon_{2})$ parameter plane in Fig. \ref{rossler}(a). As perceived, increasing $\epsilon_{2}$ reduces the critical values of $\epsilon_{1}$ to achieve the synchrony. By introducing $\epsilon_{SI}=0.00005$, a significant enhancement in synchronization region is obtained in Fig. \ref{rossler}(b). Thereafter, we investigate the complete synchrony in $(\epsilon_{1},\epsilon_{SI})$ and $(\epsilon_{2},\epsilon_{SI})$ parameter planes with fixed values of $\epsilon_{2}=0.005$ and $\epsilon_{1}=0.05$, respectively. Figures \ref{rossler}(c) and \ref{rossler}(d) depict the corresponding results in terms of $E$. It is certainly observable that the critical values of $\epsilon_{1}$ and $\epsilon_{2}$ decrease with increasing higher-order coupling strength $\epsilon_{SI}$. {\it Therefore, our results indicate the enhancement in complete synchrony in the simplicial complex of R\"{o}ssler oscillators with the inclusion of higher-order nonlinear coupling mechanism.} We validate the numerical results with the theoretical conjecture obtained from Eq. \eqref{stability_9} in terms of the maximum Lyapunov exponent, similarly to the previous example of the neuronal system. Thus, the decoupled master stability equation can be written as,
\begin{equation} \label{ross_trans}
	\begin{array}{l}
		\delta\dot{x}_{i}= -\delta y_{i} -\delta z_{i}+ \epsilon_{1}\bigg[k_{1} \big\{2(\alpha-x_{0})(x_{0}-\beta)-(x_{0}-\beta)^{2} \big \} \\
		~~~~~~~~~~~~~~~~~~~~~~      -2\gamma^{[1]}_{i}(\alpha-x_{0})(x_{0}-\beta)\bigg]\delta x_{i},\\
		
		\delta \dot{y}_{i}= \delta x_{i} +a \delta y_{i}+ \epsilon_{2}\bigg[k_{2} \big\{2(\alpha-y_{0})(y_{0}-\beta)-(y_{0}-\beta)^{2} \big\} \\
		 -2\gamma^{[2]}_{i}(\alpha-y_{0})(y_{0}-\beta) \bigg]\delta y_{i}+ \epsilon_{SI} \bigg[2k^{[SI]}\{2(\alpha-y_{0})(y_{0}-\beta) 
		\\ -(y_{0}-\beta)^{2}\} -4\gamma^{[SI]}_{i}(\alpha-y_{0})(y_{0}-\beta) \bigg]\delta y_{i}, \\\\
		
		\delta \dot{z}_{i}= z_{0}\delta x_{i} +(x_{0}-c)\delta z_{i}, ~~~~~~ i=2,3,\cdots,N, 
	\end{array}
\end{equation}
where the eigenvalues of the Laplacian $\mathscr{L}^{[1]}$ are symbolized by $\gamma^{[1]}_{1}(=0) < \gamma^{[1]}_{2} \leq \cdots \leq \gamma^{[1]}_{N}$. $\gamma^{[2]}_{i}=N$ and $\gamma^{[SI]}_{i}=N(N-2)$, $i=2,3,\cdots,N$ are the eigenvalues of the Laplacians $\mathscr{L}^{[2]}$ and $\mathscr{L}^{[SI]}$, respectively. The number of links adjacent to each node in second layer is given by $k_{2}=(N-1)$, and each node is involved in $k^{[SI]}= \begin{pmatrix} N-1 \\ 2 \end{pmatrix}$ numbers of $2$-simplices. The solid black curves imposed in Figs. \ref{rossler}(a)-\ref{rossler}(d) represent analytical synchronization thresholds obtained from Eq. \eqref{ross_trans}, solved for zero value of maximum Lyapunov exponent, which affirms that the numerical findings accord well with our theoretical predictions.      
\section{Conclusion} \label{conclusion}
To conclude, here we have developed the most generic model that accounts for higher-order interactions between dynamical units in simplicial complexes with multiple interaction layers and studied the complete synchronization phenomena. In the presence of arbitrary sort of coupling schemes, we derive the invariance condition for the synchronization state. Under the requirements of invariance condition, we determine the necessary condition for the stable synchronization state, which certainly generalizes the well-known MSF approach \cite{msf} to multilayer structures of simplicial complexes. The intricacy of multiple layers and arbitrary higher-order coupling in the considered system is reflected by the fact that our formalism yields a set of coupled linear differential equations (Eq. \eqref{stability_8b}) instead of a single parametric variational equation. However, we have shown that, for a certain instance, our formalism gives a set of uncoupled parametric variational equations (Eq. \eqref{stability_9}) with dimensions equal to the dimension of a single dynamical unit. Finally, a set of numerical results have been added to our theoretical derivations, confirming the legitimacy and applicability of the methodology.
\par  Our investigation is based on the most prevalent and well-studied synchronization behavior, namely, the complete synchronization phenomenon. However, many additional types of synchrony occurs in the systems with multiple interaction layers such as chimeras \cite{chimera1}, cluster synchronization \cite{cluster1}, relay synchronization \cite{relay2}, etc. All of these synchronization states have been investigated in multilayer systems with solely pairwise interactions. The formation of such states, or indeed fresh ones, in multilayered systems with higher-order interactions, along with their stability, are both fascinating challenges that provide way to future investigation. Apart from that, our work can be extended to higher-order structures with multiple interaction layers having directed connection topology, which has very recently been investigated for mono-layer systems with higher-order interactions \cite{directed_higher}.  Furthermore, we have focused only on the situation of clique complexes, it is still unclear how our numerical conclusions (e.g., enhancement of synchronous region) would alter if we took into account other higher-order structures, such as hypergraphs, and more generalized simplicial complexes.

\bibliographystyle{apsrev4-1} 
%

\end{document}